\documentstyle[pra,aps,epsf]{revtex}

\newcommand{\nn}{\noindent}
\newcommand{\beq}{\begin{equation}}
\newcommand{\eeq}{\end{equation}}
\newcommand{\bes}{\begin{eqnarray}}
\newcommand{\ees}{\end{eqnarray}}

\begin{document}
\draft
\thispagestyle{empty}
\title{
Possible violation of Newtonian gravitational law at small 
distances and constraints on it from the Casimir effect
}
\author{
V.~M.~Mostepanenko\footnote{On leave from A.Friedmann Laboratory
for Theoretical Physics, St.Petersburg, Russia.
\\Electronic  address:  mostep@lafex.cbpf.br}, 
M.~Novello\footnote{Electronic  address: novello@lafex.cbpf.br}
}

\address
{
Centro Brasileiro de Pesquisas F\'{\i}sicas, Rua Dr.~Xavier Sigaud, 150\\
Urca 22290--180, Rio de Janeiro, RJ --- Brazil
}

\maketitle
\begin{abstract}
The recent ideas that the gravitational and gauge interactions become
united at the weak scale lead to Yukawa-type corrections to the
Newtonian gravitational law at small distances. We briefly summarize
the best constraints on these corrections obtained recently from the
experiments on the measurement of the Casimir force. The new
constraints on the Yukawa-type interaction are derived from the
latest Casimir force measurement between a large gold coated sphere
and flat disk using an atomic force microscope. The obtained
constraints are stronger up to 19 times comparing the previous experiment
with aluminum surfaces and up to 4500 times comparing the Casimir
force measurements between dielectrics. The application range of
constraints obtained by means of an atomic force microscope is
extended.
\end{abstract}

\pacs{14.80.--j, 04.65.+e, 11.30.Pb, 12.20.Fv}

{
\section{Introduction}

It is common knowledge that the gravitational interaction holds a unique
position with respect to the other fundamental interactions described
by the Standard Model. Up to now, there is no renormalized Quantum
Gravity and the obstacles placed on the way to such a theory seem to
be insurmountable. Gravitational physics suffers also from a poor
experimental test. Even the classical Newtonian gravitational law which
is reliably proved at large distances is lacking experimental
confirmation at the distances of order 1\,mm or less.
At the same time it is generally believed to be correct up to the Planck
distances $\sqrt{G\hbar/c^3}\sim 10^{-35}\,$m. Needless to say that it
is a far-ranging extrapolation over 33 orders of magnitude.

Corrections to Newtonian gravitational law at small distances are
predicted by unified gauge theories, supersymmetry, supergravity, and
string theory. They are mediated by light and massless elementary
particles like scalar axion, graviphoton, dilaton, arion, and others.
The exchange of these particles between atoms leads to an interatomic
interaction described by Yukawa- or power-type effective potentials.
Constraints for their parameters (interaction constants $\alpha$,
$\lambda_n$, and interaction range $\lambda$ in case of a massive
particle) is the subject for considerable study (see the monograph
\cite{1} and references therein). The gravitational experiments of
Cavendish- and E\"{o}tvos-type lead to rather strong constraints over 
a distanse range $10^{-2}\,\mbox{m}<\lambda<10^6\,$km \cite{2}.
At submillimeter range the existence of corrections to Newtonian
gravity is not excluded experimentally which are in excess of it
by many orders.

The only constraints on non-Newtonian gravity at a submillimeter range
follow from the measurements of the van der Waals and Casimir force
(see, e.g., \cite{3,4,5}). Until recently, they were not enough 
restrictive in spite of the quick progress in experimental technique
and increased accuracy of force measurements. In \cite{6} the Casimir
force between the metallic surfaces of a disk and a spherical lens
was measured by the use of torsion pendulum. The obtained experimental
results and the extent of their agreement with theory were used in
\cite{7} to obtain stronger constraints on the corrections to
Newtonian gravity in a submillimeter range. The strengthening up to
30 times comparing the previously known constraints was obtained within
the range
$2.2\times 10^{-7}\,\mbox{m}\leq\lambda\leq 1.6\times 10^{-4}\,$m
(see also \cite{8}). In \cite{9,10,11} the results of the Casimir
force measurements between a metallized disk and a sphere attached to 
a cantilever of an atomic force microscope were reported and accurately
confronted with a theory. They were used in \cite{12,13} to constrain
the non-Newtonian gravity. The strengthening of constraints up to 560
times comparing the former Casimir force measurements between dielectrics 
was obtained in the interaction range
$5.9\times 10^{-9}\,\mbox{m}\leq\lambda\leq 1.15\times 10^{-7}\,$m.

In the present paper we discuss the possible violation of Newtonian
gravitational law at small distances and constraints on it following from
the Casimir effect. In Sec.~2 the recent impressive ideas are considered 
that the gravitational and gauge interactions become united at the
weak scale (see, e.g., \cite{14}). These ideas in the context of string
theory inevitably lead to the existence of Yukawa-type corrections to the
Newtonian gravitational law at moderate distances in addition to the
well known arguments in support of such corrections presented above.
Extra spatial dimensions cause even more drastic change of gravitational 
law at small distances. In Sec.~3 the constraints on the parameters
of Yukawa-type interactions are discussed which were obtained recently
from the Casimir force measurements of papers \cite{9,10,11} between
gold and aliminum surfaces.  In Sec.~4 the new constraints on the Yukawa
interaction are obtained following from the latest Casimir force
measurement between gold surfaces by means of an atomic force
microscope (these experimental results can be found in \cite{15}).
In Sec.~5 (conclusions and discussion) the importance of the new
Casimir force measurements for the elementary particle physics,
astrophysics and cosmology is underlined. 
Also the prospects
for further strengthening of constraints  on non-Newtonian gravity
in the submillimeter range are outlined.

Below we use units in which $\hbar=c=1$.

\section{The Yukawa-type corrections to Newtonian gravity}

As was told in the introduction, the corrections to Newtonian gravitational
law are predicted by the unified gauge theories, supersymmetry,
supergravity, and string theory. The effective potential of gravitational
interaction between two atoms with account of such corrections can be
represented in the form
\beq
V(r_{12})=-\frac{GM_1M_2}{r_{12}}\left(
1+\alpha_Ge^{-r_{12}/\lambda}\right),
\label{1}
\eeq
\nn
where $M_{1,2}$ are the masses of the atoms, $r_{12}$ is the distance
between them, $G$ is Newtonian gravitational constant, $\alpha_G$ is
a dimensionless interaction constant, $\lambda$ is the interaction range.
In the case that the Yukawa-type interaction is mediated by a light
particle of mass $m$ the interaction range is described by the Compton
wave length of this particle, so that $\lambda=1/m$.

According to recent ideas the gravitational and gauge interactions may
become united at the weak scale $F\sim 1\,$TeV$=10^3\,$GeV, and
the weakness of gravity at macroscopic distances is explained by the
existence of $n\geq 2$ compact (but rather large) extra spatial
dimensions \cite{14,16,17,18}. The consequence of these ideas is
that the gravitational interaction is described by Eq.~(\ref{1})
for the distances much larger than the size of characteristic
compactification dimension \cite{19,20,21}. One can arrive to
potential (\ref{1}) as follows.

Let $n$ extra dimensions be compactified by making a periodic
identification with a period $R$. If one mass $M_1$ is placed at the
origin and the test mass $M_2$ is at the distance
$r_{12}\ll R$ from $M_1$ the force law of $N=(4+n)$-dimensional 
space-time is
\beq
F_N(r_{12})=-G_N\frac{M_1M_2}{r_{12}^{N-2}},
\label{2}
\eeq
\nn
which provides the continuity of force lines in $(N-1)$-dimensional
space. If $r_{12}\gg R$ one obtains the usual Newtonian force
 \beq
F(r_{12})=-G\frac{M_1M_2}{r_{12}^{2}},
\label{3}
\eeq
\nn
where the Gauss law can be used to find the connection between the
$N$-dimensional and the usual Newtonian gravitational constants \cite{19}
\beq
G=\frac{\pi^{\frac{N-3}{2}}}{2\Gamma\left(\frac{N-1}{2}\right)}
\frac{G_N}{R^{N-4}}.
\label{4}
\eeq
\nn
If there is no extra dimensions ($n=0$, $N=4$) one finds from (\ref{4})
$G=G_N$ as it should be. Taking into account the connection between
the gravitational constants $G,\,G_N$ and the respective Planckean
energy scales
\beq
M_{Pl}^2=\frac{1}{G},\qquad
M_{Pl,\,N}^{N-2}=
\frac{2\Gamma\left(\frac{N-1}{2}\right)}{G_N\pi^{\frac{N-3}{2}}},
\label{5}
\eeq
\nn
Eq.~(\ref{4}) turns out to be equivalent to
\beq
M_{Pl}^2=M_{Pl,\,N}^{N-2}\,R^{N-4}.
\label{6}
\eeq

This result gives the possibility to estimate the allowed values of the
compactification dimension $R$ and the required number of extra
dimensions. Putting $N$-dimensional Planckean scale be equal to the
weak scale, i.e. $M_{Pl,\,N}=F$, one obtains from (\ref{6}) \cite{14}
\beq
R=\frac{1}{F}\left(\frac{M_{Pl}}{F}\right)^{\frac{2}{N-4}}\sim
10^{\frac{30}{N-4}}\frac{1}{F}\sim
10^{\frac{30}{N-4}-17}\,\mbox{cm}
\label{7}
\eeq
\nn
(we remind that $M_{Pl}\approx 2.4\times 10^{18}\,$GeV).
Evidently, one extra dimension is impermissable because for $N=5$ it
follows $R\sim 10^{13}\,$cm which is in contradiction with the confirmed
validity of Newtonian gravitational law at the scales of solar system.
But already $n=2$ ($N=6$) leads to $R\sim 10^{-2}\,$cm which is
permitted by the results of gravitational measurements.

The above considerations show that the gravitational law may vary
from (\ref{2}) at small, submillimeter distances to the usual form
(\ref{3}) at relatively large distances. The corrections to 
Eqs.~(\ref{2}), (\ref{3}) can be found by considering the Newtonian
limit of $N$-dimensional Einstein gravity \cite{19,20,21}. At the
distances $r_{12}\ll R$ the corrections to Eq.~(\ref{2}) are of
power type. The corrections at the distances $r_{12}\gg R$ hold the
greatest interest because they can be tested experimentally. It is
significant that they are of Yukawa-type, so that remote from the
compactification scale the gravitational potential is given by
Eq.~(\ref{1}). Coefficient $\alpha_G$ of Eq.~(\ref{1}) depends on
compactification geometry and on the number of extra dimensions.
By way of example, if $n$ extra dimensions have the topology of
$n$-torus $\alpha_G^{(n)}=2n$, and if they have the topology of
$n$-sphere $\alpha_G^{(n)}=n+1$ \cite{21}.

The above models with $n=6$ can be formulated within type I or type
IIB string theories \cite{16,18} (in the case of M theory $n=7$).
In so doing, gravitons are described by closed strings and propagate
in $N$-dimensional bulk. The particles of the Standard Model are
described by open strings living on $(3+1)$-dimensional wall. This
wall should have a thickness of order $F^{-1}\sim 10^{-17}\,$cm in the
extra dimensions. Gravity becomes unified with the gauge interactions
of Standard Model at the weak scale $F$. The usual Newtonian
gravitational constant $G$ loses its status of a fundamental
constant. It is a multidimensional gravitational constant $G_N$
which acquires a meaning of the fundamental one.  Note that the
separated character of gravitons which may propagate freely in extra
dimensions, while all ordinary particles cannot do so, is in some
analogy with the field theory of gravity \cite{22} where the gravity
to gravity interaction is quite distinct from the interaction of
matter to gravity.

The possibility of serious variations of the gravitational law
in submillimeter range makes much-needed the performance of new
experiments. As was stressed, e.g., in \cite{3,5}, the Casimir
effect may well become a new method for experimental verification
of fundamental physical theories. In \cite{23} the measurements
of the Casimir force between plane plates were considered in
order to restrict the extra dimensions and string-inspired forces.
In the next two sections the recent experiments on measuring the
Casimir force between a disk and a spherical lens (sphere) are
used for the same purpose and new more strong constraints are
obtained.

\section{Review of the best constraints on Yukawa-type interaction
in submillimeter range}

It has been known that the Casimir and van der Waals force
measurements between dielectrics (see, e.g, \cite{24}) lead
to the strongest constraints on the constants of Yukawa-type
interaction given by the second term of Eq.~(\ref{1}) with a range
of action $10^{-9}\,\mbox{m}<\lambda<10^{-4}\,$m \cite{3,4}.
In \cite{6} the Casimir force between two metallized surfaces
of a flat disk and a spherical lens was measured with the use of
torsion pendulum. The outer metallic layer of gold, covering the test
bodies, had the thickness of 0.5\,$\mu$m. The absolute error of the
force measurements in \cite{6} was $\Delta F=10^{-11}\,$N for the
distances $a$ between the disk and the lens in the range
$1\,\mu\mbox{m}\leq a\leq 6\,\mu$m. In the limits of this error
the theoretical expression for the Casimir force
\beq
F^{(0)}(a)=-\frac{\pi^3}{360}\frac{R}{a^3}
\label{8}
\eeq
\nn
was confirmed (where $R$ is lens curvature radius).

No corrections to Eq.~(\ref{8}) due to surface roughness,
finite conductivity of the boundary metal
or nonzero temperature
was recorded. These corrections, however, 
may not lie in the limits of the absolute error $\Delta F$. By way of 
example, at $a\approx 1\,\mu$m the roughness correction
may be around 12\% of $F^{(0)}$ or even larger
\cite{7}, and the finite conductivity correction 
for the gold surfaces at $1\,\mu$m separation is 10\% of
$F^{(0)}$ \cite{25}. 
(Remind that $\Delta F$ is around 3\% of $F^{(0)}$ at $a=1\,\mu$m.) 
As to the temperature correction, 
it achieves 174\% of $F^{(0)}$ at the separation
$a=6\,\mu$m, where, however, $\Delta F$ is around 700\% of $F^{(0)}$. 
By this reason the constraints for Yukawa-type
interaction following from the experiment \cite{6} were found
from the inequality \cite{7}
\beq
|F_{th}(a)-F^{(0)}(a)|\leq\Delta F,
\label{9}
\eeq
\nn
where $F_{th}$ is the theoretical force value, including $F^{(0)}$,
all the corrections to it mentioned above, and also the hypothetical
Yukawa-type interaction calculated in \cite{7} at experimental
configuration (remind that the sign of finite conductivity
correction is opposite to the signs of other corrections).

The obtained results are shown in Fig.~1 where the regions above
the curves are prohibited and below the curves are permitted.
Curve 1 represents the constraints following from the Cavendish-type
experiment of Ref.~\cite{26}, curves 2,a and 2,b are calculated
by Eq.~(\ref{9}) for $\alpha_G>0$ respectively $\alpha_G<0$ \cite{7}.
Curve 3 was obtained earlier (see, e.g., \cite{3}) by the results of
Casimir force measurements between dielectrics. The strengthening of
constraints given by the curves 2,a and 2,b comparing the curve 3
is up to a factor 30 in the interaction range
$2.2\times 10^{-7}\,\mbox{m}\leq\lambda\leq 1.6\times 10^{-4}\,$m
(a bit different result was obtained later in \cite{8} where the
corrections to the ideal Casimir force (\ref{8}) were not taken
into account).

In \cite{9,10} the results of the Casimir force measurements between
a flat disk and a sphere by means of an atomic force microscope were
presented in comparison with the theory taking into account the finite 
conductivity and roughness corrections. Temperature corrections are not
essential in the interactiuon range
$0.1\,\mu\mbox{m}<a<0.9\,\mu$m of \cite{9,10}. The test bodies were
covered by the aluminum layer of 300\,nm thickness and $Au/Pd$
layer of the thickness 20\,nm (the latter is transparent for
electromagnetic oscillations of characteristic frequency). The
absolute error of the force measurements in \cite{9,10} was
$\Delta F=2\times 10^{-12}\,$N. In the limits of this error the
theoretical expression for the Casimir force with corrections to it due
to both surface roughness and finite conductivity was confirmed. 
The theoretical expression for the Yukawa-type interaction 
in experimental configuration of \cite{9,10} was obtained in \cite{12}.
The constraints on the parameters $\alpha_G,\,\lambda$
calculated in \cite{12} from the inequality
\beq
|F_{Yu}(a)|\leq\Delta F,
\label{10}
\eeq
\nn
turned out to be the best ones in the interaction range
$5.9\times 10^{-9}\,\mbox{m}\leq\lambda\leq 10^{-7}\,$m.
They are stronger up to 140 times than the previously known ones from
the Casimir force measurements between dielectrics (note that here all 
the corrections were included into the force under measuring; by this
reason the constraints on $|\alpha_G|$ rather than on $\alpha_G$
were obtained).

In \cite{11} the improved precision measurement of the Casimir force
was performed by means of an atomic force microscope.
The experimental improvements which include vibration isolation, lower 
systematic errors, and independent measurement of surface separation gave 
the possibility to decrease the absolute error of force measurement by 
a factor 2. Also the smoother $Al$ coating with thickness 250\,nm
was used and thinner external $Au/Pd$ layer of the thickness 7.9\,nm.
The Yukawa-type hypothetical force in the configuration of \cite{11}
was calculated in \cite{13} where the stronger constraints on
$\alpha_G,\,\lambda$ were also obtained using the inequality (\ref{10}).
These constraints are presented in Fig.~1 (curve 4). They turned out 
to be up to four times stronger than the constraints \cite{12} obtained 
from the previous experiment \cite{9,10} within a bit wider
interaction range
$5.9\times 10^{-9}\,\mbox{m}\leq\lambda\leq 1.15\times 10^{-7}\,$m.
The total strengthening of constraints on the corrections
to Newtonian gravitational law
from the measurements of the Casimir force by means of atomic force
microscope has reached 560 times within the $\lambda$-interval
mentioned above. 

\section{New constraints on the Yukawa-type interaction from the
Casimir force measurement between gold surfaces by means of atomic 
force microscope}

Recently one more measurement of the Casimir force was performed using
the atomic force microscope \cite{15}. The test bodies (sphere and a disk)
were coated by gold instead of aluminum which removes some difficulties
connected with the additional thin $Au/Pd$ layers used in the
previous measurements \cite{9,10,11} to prevent the oxidation processes 
on $Al$ surfaces. The used polystyrene sphere coated by gold layer
was of diameter $2R=191.3\,\mu$m and a sapphire disk had a diameter
$2L=1\,$cm, and a thickness $D=1\,$mm. The thickness of the gold
coating on both test bodies was $\Delta=86.6\,$nm. This can be considered 
as an infinitely thick in relation to the Casimir force measurements.
The root mean square roughness amplitude of the gold surfaces was
decreased until 1\,nm which makes roughness corrections negligibly
small. The measurements were performed at smaller separations, i.e.
$62\,\mbox{nm}\leq a\leq 350\,$nm. The absolute error of force measurements
was, however, $\Delta F=3.5\times 10^{-12}\,$N, i.e., a bit larger
than in the previous experiments. The reason is the thinner gold
coating used in \cite{15} which led to poor termal conductivity of
the cantilever. At smaller separations of about 65\,nm this error
is less than 1\% of the measured Casimir force.

Now let us calculate the gravitational force acting in experimental
configuration due to the potential (\ref{1}). The Newtonian contribution
is found to be negligible. Actually, due to the inequality $R\ll L$
each atom of the sphere can be considered as if it would be placed
above the center of the disk. Then the vertical component of the
Newtonian gravitational force acting between the sphere atom of a mass
$M_1$ situated at a height $l\ll L$ and the disk is
\beq
f_{N,z}(l)=\frac{\partial}{\partial l}\left[
GM_1\rho2\pi
\int\limits_{0}^{L}r\,dr
\int\limits_{l}^{l+D}\frac{dz}{\sqrt{r^2+z^2}}\right]
\approx -2\pi GM_1\rho D\left[
1-\frac{D+2l}{2L}\right],
\label{11}
\eeq
\nn
where $\rho$ is the disk density, and only the first order terms in
$D/L$ and $l/L$ are retained.

The Newtonian gravitational force acting between the disk and the sphere
is obtained from (\ref{11}) by integration over the sphere volume
\beq
F_{N,z}\approx -\frac{8}{3}\pi^2G\rho\rho^{\prime}DR^3\left(
1-\frac{D}{2L}-\frac{R}{L}\right),
\label{12}
\eeq
\nn
where $\rho^{\prime}$ is the density of the sphere material.

Even with sphere and disk made of the vacuo-distilled gold as a whole
with $\rho=\rho^{\prime}=18.88\times10^3\,$kg/m${}^3$ one arrives from
(\ref{12}) to the negligibly small value of
$F_{N,z}\approx 6\times 10^{-16}\,\mbox{N}\ll\Delta F$.

The Yukawa-type addition to the Newtonian gravity which is due to the
second term of the potential (\ref{1}) should be calculated with account 
of true materials of the test bodies (for a polystyrene sphere
$\rho^{\prime}=1.06\times10^3\,$kg/m${}^3$,
for a sapphire disk $\rho=4.0\times10^3\,$kg/m${}^3$,
and for the gold covering layers
$\rho_1=18.88\times10^3\,$kg/m${}^3$).
It can be easily obtained using the same procedure which was applied 
in the case of the Newtonian gravitational force. The result is
\beq
F_{Yu}(a)=-4\pi^2G\alpha_G\lambda^3e^{-a/\lambda}R
\left[\rho_1-(\rho_1-\rho)e^{-\Delta/\lambda}
\right]\, \left[\rho_1-(\rho_1-\rho^{\prime})e^{-\Delta/\lambda}
\right].
\label{13}
\eeq

According to \cite{15} the theoretical value of the Casimir force was
confirmed within the limits of $\Delta F=3.5\times 10^{-12}\,$N and
no hypothetical force was observed. In such a situation, the
constraints on $\alpha_G$ can be obtained from the inequality
(\ref{10}). The strongest constraints follow for the smallest possible
values of $a\approx 65\,$nm. The computational results are presented
by curve 5 in Fig.~1.

As is seen from Fig.~1, the Casimir force measurement between the gold
surfaces by means of an atomic force microscope gives the possibility
to strengthen the previuosly known constraints (curve 4) up to 19 times
within a range 
$4.3\times 10^{-9}\,\mbox{m}\leq\lambda\leq 1.5\times 10^{-7}\,$m.
The largest strengthening takes place for $\lambda=$(5-10)\,nm.
Comparing the constraints obtained from the Casimir and van der
Waals force measurements between dielectrics (curves 3 and 6) the
strengthening up to 4500 was achieved by the Casimir force measurement
\cite{15} between gold surfaces using the atomic force microscope. 

\section{Conclusions and discussion}

As indicated above, there is abundant evidence that the gravitational
interaction at small distances undergo deviations from the Newtonian law.
These deviations can be described by the Yukawa-type potential.
They were predicted in the theoretical schemes with the quantum
gravity scale both of order $10^{18}\,$GeV and $10^3\,$GeV. 
In the latter case the problem of experimental search for such
deviations takes on great significance.
The existence of large extra dimensions can radically alter many
concepts of space-time, elementary particle physics, astrophysics and
cosmology. To cite an example, the highest temperature at which
the Universe was born turns out to be $\sim 10^3\,$GeV instead of
$10^{18}\,$GeV. Although this does not touch the era of big-bang
nucleosynthesis which begins at lower temperatures of about 1\,MeV the
theory of the very early Universe including inflation may be
significantly changed \cite{19}. 

The Casimir and van der Waals force measurements is the main source
of constraints on the Yukawa-type interactions at small distances.
In the present paper we have reviewed the latest advances deduced in
such a manner in the submillimeter interaction range. The new constraints
were obtained also following from the recent Casimir force
measurements between gold surfaces by means of an atomic force
microscope \cite{15}. They are found to be up to 19 times stronger
than the previously reported \cite{13} and up to 4500 stronger than
the well-known constraints obtained from the Casimir and van der
Waals force measurements between dielectrics. Although the interaction
range where the new constraints are valid was extended, there is yet a gap
(shown by curve 3 in Fig.~1) up to the curves 2,a, 2,b obtained from
the Casimir force measurement by the use of torsion pendulum \cite{6}.

As it is seen from Fig.~1, much work should be done in order to test 
experimentally predictions of the models with weak-scale compactification 
characterized by the value $\alpha_G\sim 10$ (see Sec.~2). As was
shown in \cite{7} the constraints following from the experiment
\cite{6} can be improved up to four orders of magnitude in the range
around $\lambda=10^{-4}\,$m. This is just right to attain the
desirable values of $\alpha_G$. As to the experiments using the
atomic force microscope \cite{9,10,11,15}, where the larger strengthening
of the previously known constraints is already achieved (up to 4500
times), there remains almost fifteen orders more needed to achieve
the value $\alpha_G\sim 10$ in the interaction range
$\lambda\leq 10^{-7}\,$m. Thus, it is desirable here not only to
increase the strength of constraints but also to move the interaction
range under examination to larger $\lambda$ (e.g., by the increase
of a sphere radius and the space separation to the disk). 
In conclusion it may be said that the compact and relatively cheap 
laboratory experiments on the measurement of the Casimir force offer
important advantages over the commonly used techniques intended for the
investigation of fundamental interactions and new elementary
particles.

\section*{Acknowledgments}

The authors are grateful to G.~L.~Klimchitskaya, D.~E.~Krause, and
U.~Mohideen for helpful discussions. V.M.M. is grateful to the
Centro Brasileiro de Pesquisas F\'{\i}sicas where the work was
performed for kind hospitality. He acknowledges the financial
support from FAPERJ. M.N. was partly supported by CNPq.

}
\newpage
\begin{figure}[h]
\epsfxsize=15cm\centerline{\epsffile{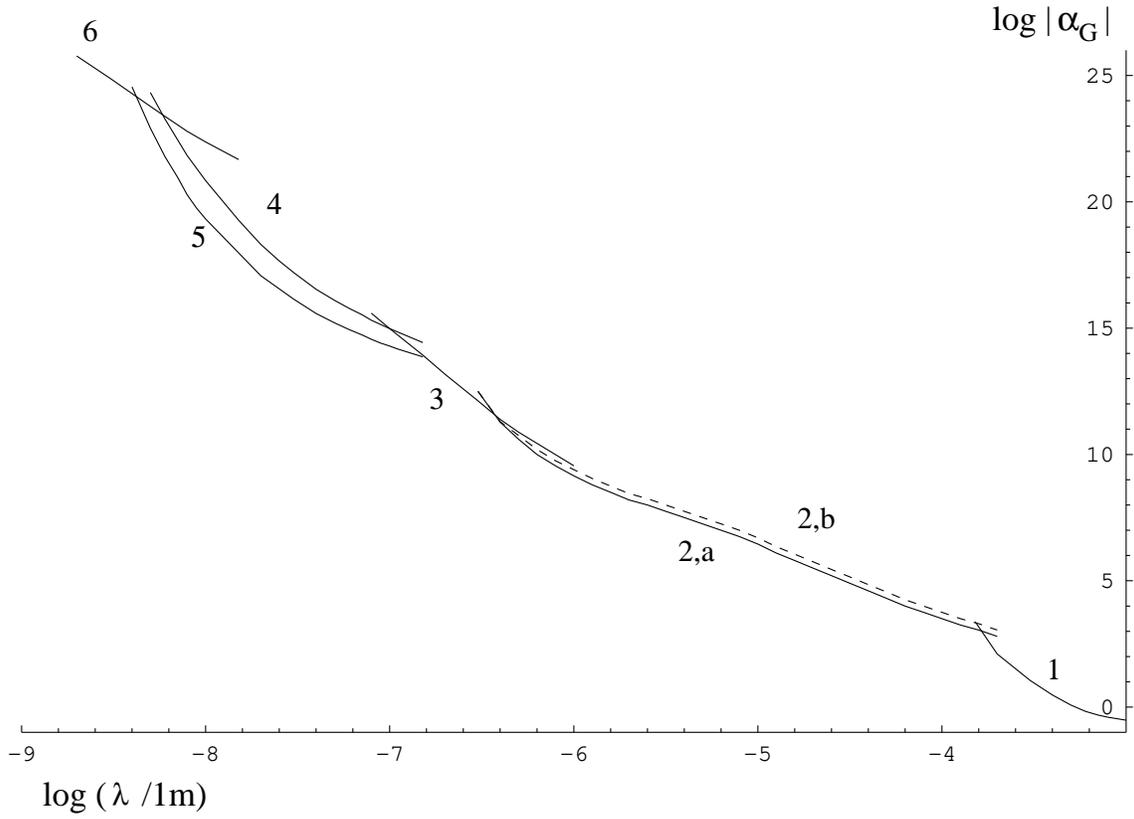} }
\vspace{0.5cm}
\caption{
Constraints on the Yukawa-type  interaction.
Curve 1 follows from the Cavendish-type experiment of Ref.~[26].
Curves 2,a, 2,b follow from the Casimir force measurement by means
of a torsion pendulum [6]. 
Curves 3 and 6 were obtained
from the Casimir and van der Waals force measurements  between dielectrics.
Curve 4 follows [13] from the  Casimir force 
 measurement  between aluminum surfaces by means of
an atomic force microscope [11], and
curve 5 is obtained in this paper from the experiment of Ref.~[15]
with gold surfaces.
The regions below the curves are permitted, and those above the curves are
prohibited.
}
\end{figure}
\end{document}